\expandafter \def \csname CHAPLABELintro\endcsname {1}
\expandafter \def \csname CHAPLABELvect\endcsname {2}
\expandafter \def \csname EQLABELKpotx\endcsname {2.1?}
\expandafter \def \csname EQLABELKpotz\endcsname {2.2?}
\expandafter \def \csname EQLABELhetf\endcsname {2.3?}
\expandafter \def \csname EQLABELftree\endcsname {2.4?}
\expandafter \def \csname EQLABELKtree\endcsname {2.5?}
\expandafter \def \csname EQLABELtreemetric\endcsname {2.6?}
\expandafter \def \csname EQLABELfdual\endcsname {2.7?}
\expandafter \def \csname EQLABELnH\endcsname {2.8?}
\expandafter \def \csname TABLABELweights\endcsname {2.1?}
\expandafter \def \csname EQLABELfcubic\endcsname {2.9?}
\expandafter \def \csname CHAPLABELcmap\endcsname {3}
\expandafter \def \csname EQLABELcmapgen\endcsname {3.1?}
\expandafter \def \csname EQLABELsnmapgen\endcsname {3.2?}
\expandafter \def \csname EQLABELduquat\endcsname {3.3?}
\expandafter \def \csname EQLABELefL\endcsname {3.4?}
\expandafter \def \csname EQLABELmodsred\endcsname {3.5?}
\expandafter \def \csname TABLABELcmapsym\endcsname {3.1?}
\expandafter \def \csname EQLABELvmulL\endcsname {3.6?}
\expandafter \def \csname EQLABELdredL\endcsname {3.7?}
\expandafter \def \csname EQLABELdredLm\endcsname {3.8?}
\expandafter \def \csname CHAPLABELhet\endcsname {4}
\expandafter \def \csname EQLABELinters\endcsname {4.1?}
\expandafter \def \csname EQLABELmodK\endcsname {4.2?}
\expandafter \def \csname EQLABELmodKS\endcsname {4.3?}
\expandafter \def \csname EQLABELmodKo\endcsname {4.4?}
\expandafter \def \csname EQLABELmodKSo\endcsname {4.5?}
\expandafter \def \csname EQLABELftmir\endcsname {4.6?}
\expandafter \def \csname EQLABELMvtmir\endcsname {4.7?}
\expandafter \def \csname EQLABELexprepiia\endcsname {4.8?}
\expandafter \def \csname EQLABELmap\endcsname {4.9?}
\expandafter \def \csname EQLABELexprephet\endcsname {4.10?}
\expandafter \def \csname TABLABELinstantons\endcsname {4.1?}
\expandafter \def \csname EQLABELexprep\endcsname {4.11?}
\expandafter \def \csname CHAPLABELdisc\endcsname {5}

\font\eightrm=cmr8 at 8pt

\font\seventeenrm=cmr17 at 17pt
\font\twentyonerm=cmr17 at 21pt

\font\ss=cmss10

\font\csc=cmcsc10

\font\twelvecal=cmsy10 at 12pt

\font\twelvemath=cmmi12

\font\seventeenbold=cmbx7 at 17pt

\font\fively=lasy5
\font\sevenly=lasy7
\font\tenly=lasy10

\textfont10=\tenly
\scriptfont10=\sevenly
\scriptscriptfont10=\fively
\magnification=1200
\parskip=10pt
\parindent=20pt
\def\today{\ifcase\month\or January\or February\or March\or April\or May\or
June
       \or July\or August\or September\or October\or November\or December\fi
       \space\number\day, \number\year}

\def\title#1{\footline={\ifnum\pageno<2\hfil
       \else\hss\tenrm\folio\hss\fi}\vskip1truein\centerline{{#1}
       \footnote{\raise1ex\hbox{*}}{\eightrm Supported in part
       by the Robert A. Welch Foundation and N.S.F. Grants
       PHY-880637 and\break PHY-8605978.}}}

\def\newpage{\vfill\eject}
\def\abstract#1{\centerline{\bf ABSTRACT}\vskip.2truein{\narrower\noindent#1
       \smallskip}}

\def\runninghead#1#2{\voffset=2\baselineskip\nopagenumbers
       \headline={\ifodd\pageno\rightheadline\else \leftheadline\fi}
       \def\rightheadline{{\sl#1}\hfill{\rm\folio}}
       \def\leftheadline{{\rm\folio}\hfill{\sl#2}}}
\def\SS{\mathhexbox278}

\newcount\footnoteno
\def\Footnote#1{\advance\footnoteno by 1
                \let\SF=\empty
                \ifhmode\edef\SF{\spacefactor=\the\spacefactor}\/\fi
                $^{\the\footnoteno}$\ignorespaces
                \SF\vfootnote{$^{\the\footnoteno}$}{#1}}

\def\place#1#2#3{\vbox to0pt{\kern-\parskip\kern-7pt
                             \kern-#2truein\hbox{\kern#1truein #3}
                             \vss}\nointerlineskip}
\def\figurecaption#1#2{\kern.75truein\vbox{\hsize=5truein\noindent{\bf Figure
    \figlabel{#1}:} #2}}
\def\tablecaption#1#2{\kern.75truein\lower12truept\hbox{\vbox{\hsize=5truein
    \noindent{\bf Table\hskip5truept\tablabel{#1}:} #2}}}
\def\boxed#1{\lower3pt\hbox{
                       \vbox{\hrule\hbox{\vrule

\vbox{\kern2pt\hbox{\kern3pt#1\kern3pt}\kern3pt}\vrule}
                         \hrule}}}
\def\a{\alpha}

\def\g{\gamma}\def\G{\Gamma}

\def\l{\lambda}\def\L{\Lambda}
\def\m{\mu}
\def\n{\nu}

\def\p{\pi}

\def\S{\Sigma}

\def\O{\Omega}

\def\ca#1{\relax\ifmmode {{\cal #1}}\else $\cal #1$\fi}

\def\calb{{\cal B}}

\def\calm{{\cal M}}

\def\inbar{\vrule height1.5ex width.4pt depth0pt}
\def\IB{\relax{\rm I\kern-.18em B}}
\def\IC{\relax\hbox{\kern.25em$\inbar\kern-.3em{\rm C}$}}
\def\ID{\relax{\rm I\kern-.18em D}}
\def\IE{\relax{\rm I\kern-.18em E}}
\def\IF{\relax{\rm I\kern-.18em F}}
\def\IG{\relax\hbox{\kern.25em$\inbar\kern-.3em{\rm G}$}}
\def\IH{\relax{\rm I\kern-.18em H}}
\def\II{\relax{\rm I\kern-.18em I}}
\def\IK{\relax{\rm I\kern-.18em K}}
\def\IL{\relax{\rm I\kern-.18em L}}
\def\IM{\relax{\rm I\kern-.18em M}}
\def\IN{\relax{\rm I\kern-.18em N}}
\def\IO{\relax\hbox{\kern.25em$\inbar\kern-.3em{\rm O}$}}
\def\IP{\relax{\rm I\kern-.18em P}}
\def\IQ{\relax\hbox{\kern.25em$\inbar\kern-.3em{\rm Q}$}}
\def\IR{\relax{\rm I\kern-.18em R}}
\def\IZ{\relax\ifmmode\hbox{\ss Z\kern-.4em Z}\else{\ss Z\kern-.4em Z}\fi}
\def\IGa{\relax{\rm I}\kern-.18em\Gamma}
\def\IPi{\relax{\rm I}\kern-.18em\Pi}
\def\ITh{\relax\hbox{\kern.25em$\inbar\kern-.3em\Theta$}}
\def\IOm{\relax\thinspace\inbar\kern1.95pt\inbar\kern-5.525pt\Omega}


\def\ie{{\it i.e.,\ \/}}
\def\eg{{\it e.g.,\ \/}}

\def\cym{Calabi--Yau manifold}

\def\cyt{Calabi--Yau threefold}

\def\K{K\"ahler}

\def\H#1#2{\relax\ifmmode {H^{#1#2}}\else $H^{#1 #2}$\fi}
\def\M{\relax\ifmmode{\calm}\else $\calm$\fi}

\def\Bigcheck{\lower3.8pt\hbox{\smash{\hbox{{\twentyonerm \v{}}}}}}
\def\bigboldcheck{\smash{\hbox{{\seventeenbold\v{}}}}}

\def\Bighat{\lower3.8pt\hbox{\smash{\hbox{{\twentyonerm \^{}}}}}}

\def\Msharp{\relax\ifmmode{\calm^\sharp}\else $\smash{\calm^\sharp}$\fi}
\def\Mflat{\relax\ifmmode{\calm^\flat}\else $\smash{\calm^\flat}$\fi}
\def\preMcheck{\kern2pt\hbox{\Bigcheck\kern-12pt{$\cal M$}}}
\def\Mcheck{\relax\ifmmode\preMcheck\else $\preMcheck$\fi}
\def\preMhat{\kern2pt\hbox{\Bighat\kern-12pt{$\cal M$}}}
\def\Mhat{\relax\ifmmode\preMhat\else $\preMhat$\fi}

\def\Bsharp{\relax\ifmmode{\calb^\sharp}\else $\calb^\sharp$\fi}
\def\Bflat{\relax\ifmmode{\calb^\flat}\else $\calb^\flat$ \fi}
\def\preBcheck{\hbox{\Bigcheck\kern-9pt{$\cal B$}}}
\def\Bcheck{\relax\ifmmode\preBcheck\else $\preBcheck$\fi}
\def\preBhat{\hbox{\Bighat\kern-9pt{$\cal B$}}}
\def\Bhat{\relax\ifmmode\preBhat\else $\preBhat$\fi}

\def\figBcheck{\kern3pt\hbox{\raise1pt\hbox{\bigboldcheck}\kern-11pt
    {\twelvecal B}}}
\def\figBsharp{{\twelvecal B}\raise5pt\hbox{$\twelvemath\sharp$}}
\def\figBflat{{\twelvecal B}\raise5pt\hbox{$\twelvemath\flat$}}

\def\gcheck{\hbox{\lower2.5pt\hbox{\Bigcheck}\kern-8pt$\g$}}
\def\lhat{\hbox{\raise.5pt\hbox{\Bighat}\kern-8pt$\l$}}

\def\Fcheck{\kern2pt\hbox{\raise1pt\hbox{\Bigcheck}\kern-10pt{$\cal F$}}}
\def\Fhat{\kern2pt\hbox{\raise1pt\hbox{\Bighat}\kern-10pt{$\cal F$}}}

\def\cp#1{\relax\ifmmode {\IP\kern-2pt{}_{#1}}\else $\IP\kern-2pt{}_{#1}$\fi}
\def\h#1#2{\relax\ifmmode {b_{#1#2}}\else $b_{#1#2}$\fi}

\def\frac#1#2{{#1\over #2}}

\def\cone{\relax\thinspace\hbox{$<\kern-.8em{)}$}}
\mathchardef\mho"0A30

\def\-{\hphantom{-}}


\def\npb#1{Nucl.\ Phys.\ {\bf B#1}}

\def\cmp#1{Commun. Math. Phys. {\bf #1}}
\def\plb#1{Phys. Lett. {\bf #1B}}


\def\picture #1 by #2 (#3){\vbox to #2{\hrule width #1 height 0pt depth 0pt
                                       \vfill\special{picture #3}}}
\def\scaledpicture #1 by #2 (#3 scaled #4){{\dimen0=#1 \dimen1=#2
           \divide\dimen0 by 1000 \multiply\dimen0 by #4
            \divide\dimen1 by 1000 \multiply\dimen1 by #4
            \picture \dimen0 by \dimen1 (#3 scaled #4)}}
\def\illustration #1 by #2 (#3){\vbox to #2{\hrule width #1 height 0pt depth
0pt
                                       \vfill\special{illustration #3}}}
\def\scaledillustration #1 by #2 (#3 scaled #4){{\dimen0=#1 \dimen1=#2
           \divide\dimen0 by 1000 \multiply\dimen0 by #4
            \divide\dimen1 by 1000 \multiply\dimen1 by #4
            \illustration \dimen0 by \dimen1 (#3 scaled #4)}}


\def\delaOssa{\nobreak\vskip1truein\hbox to\hsize
       {\hskip 4truein Xenia de la Ossa\hfill}}

\def\hoy{\number\day\space de \ifcase\month\or enero\or febrero\or marzo\or
       abril\or mayo\or junio\or julio\or agosto\or septiembre\or octubre\or
       noviembre\or diciembre\fi\space de \number\year}


\newif\ifproofmode
\proofmodefalse

\newif\ifforwardreference
\forwardreferencefalse

\newif\ifchapternumbers
\chapternumbersfalse

\newif\ifcontinuousnumbering
\continuousnumberingfalse

\newif\iffigurechapternumbers
\figurechapternumbersfalse

\newif\ifcontinuousfigurenumbering
\continuousfigurenumberingfalse

\newif\iftablechapternumbers
\tablechapternumbersfalse

\newif\ifcontinuoustablenumbering
\continuoustablenumberingfalse

\font\eqsixrm=cmr6

\def\marginstyle{\eqsixrm}

\newtoks\chapletter
\newcount\chapno
\newcount\eqlabelno
\newcount\figureno
\newcount\tableno

\chapno=0
\eqlabelno=0
\figureno=0
\tableno=0

\def\chapfolio{\ifnum\chapno>0 \the\chapno\else\the\chapletter\fi}

\def\bumpchapno{\ifnum\chapno>-1 \global\advance\chapno by 1
\else\global\advance\chapno by -1 \setletter\chapno\fi
\ifcontinuousnumbering\else\global\eqlabelno=0 \fi
\ifcontinuousfigurenumbering\else\global\figureno=0 \fi
\ifcontinuoustablenumbering\else\global\tableno=0 \fi}

\def\setletter#1{\ifcase-#1{}\or{}%
\or\global\chapletter={A}%
\or\global\chapletter={B}%
\or\global\chapletter={C}%
\or\global\chapletter={D}%
\or\global\chapletter={E}%
\or\global\chapletter={F}%
\or\global\chapletter={G}%
\or\global\chapletter={H}%
\or\global\chapletter={I}%
\or\global\chapletter={J}%
\or\global\chapletter={K}%
\or\global\chapletter={L}%
\or\global\chapletter={M}%
\or\global\chapletter={N}%
\or\global\chapletter={O}%
\or\global\chapletter={P}%
\or\global\chapletter={Q}%
\or\global\chapletter={R}%
\or\global\chapletter={S}%
\or\global\chapletter={T}%
\or\global\chapletter={U}%
\or\global\chapletter={V}%
\or\global\chapletter={W}%
\or\global\chapletter={X}%
\or\global\chapletter={Y}%
\or\global\chapletter={Z}\fi}

\def\tempsetletter#1{\ifcase-#1{}\or{}%
\or\global\chapletter={A}%
\or\global\chapletter={B}%
\or\global\chapletter={C}%
\or\global\chapletter={D}%
\or\global\chapletter={E}%
\or\global\chapletter={F}%
\or\global\chapletter={G}%
\or\global\chapletter={H}%
\or\global\chapletter={I}%
\or\global\chapletter={J}%
\or\global\chapletter={K}%
\or\global\chapletter={L}%
\or\global\chapletter={M}%
\or\global\chapletter={N}%
\or\global\chapletter={O}%
\or\global\chapletter={P}%
\or\global\chapletter={Q}%
\or\global\chapletter={R}%
\or\global\chapletter={S}%
\or\global\chapletter={T}%
\or\global\chapletter={U}%
\or\global\chapletter={V}%
\or\global\chapletter={W}%
\or\global\chapletter={X}%
\or\global\chapletter={Y}%
\or\global\chapletter={Z}\fi}

\def\chapshow#1{\ifnum#1>0 \relax#1%
\else{\tempsetletter{\number#1}\chapno=#1\chapfolio}\fi}

\def\chaplabel#1{\bumpchapno\ifproofmode\ifforwardreference
\immediate\write1{\noexpand\expandafter\noexpand\def
\noexpand\csname CHAPLABEL#1\endcsname{\the\chapno}}\fi\fi
\global\expandafter\edef\csname CHAPLABEL#1\endcsname
{\the\chapno}\ifproofmode\llap{\hbox{\marginstyle #1\ }}\fi\chapfolio}

\def\eqnum{\global\advance\eqlabelno by 1
\eqno(\ifchapternumbers\chapfolio.\fi\the\eqlabelno)}

\def\eqlabel#1{\global\advance\eqlabelno by 1 \ifproofmode\ifforwardreference
\immediate\write1{\noexpand\expandafter\noexpand\def
\noexpand\csname EQLABEL#1\endcsname{\the\chapno.\the\eqlabelno?}}\fi\fi
\global\expandafter\edef\csname EQLABEL#1\endcsname
{\the\chapno.\the\eqlabelno?}\eqno(\ifchapternumbers\chapfolio.\fi
\the\eqlabelno)\ifproofmode\rlap{\hbox{\marginstyle #1}}\fi}

\def\eqalignnum{\global\advance\eqlabelno by 1
&(\ifchapternumbers\chapfolio.\fi\the\eqlabelno)}

\def\eqalignlabel#1{\global\advance\eqlabelno by 1 \ifproofmode
\ifforwardreference\immediate\write1{\noexpand\expandafter\noexpand\def
\noexpand\csname EQLABEL#1\endcsname{\the\chapno.\the\eqlabelno?}}\fi\fi
\global\expandafter\edef\csname EQLABEL#1\endcsname
{\the\chapno.\the\eqlabelno?}&(\ifchapternumbers\chapfolio.\fi
\the\eqlabelno)\ifproofmode\rlap{\hbox{\marginstyle #1}}\fi}

\def\eqref#1{\hbox{(\ifundefined{EQLABEL#1}***)\ifproofmode\ifforwardreference%
\else\write16{ ***Undefined Equation Reference #1*** }\fi
\else\write16{ ***Undefined Equation Reference #1*** }\fi
\else\edef\LABxx{\getlabel{EQLABEL#1}}%
\def\LAByy{\expandafter\stripchap\LABxx}\ifchapternumbers%
\chapshow{\LAByy}.\expandafter\stripeq\LABxx%
\else\ifnum\number\LAByy=\chapno\relax\expandafter\stripeq\LABxx%
\else\chapshow{\LAByy}.\expandafter\stripeq\LABxx\fi\fi)\fi}%
\ifproofmode\write2{Equation #1}\fi}

\def\fignum{\global\advance\figureno by 1
\relax\iffigurechapternumbers\chapfolio.\fi\the\figureno}

\def\figlabel#1{\global\advance\figureno by 1
\relax\ifproofmode\ifforwardreference
\immediate\write1{\noexpand\expandafter\noexpand\def
\noexpand\csname FIGLABEL#1\endcsname{\the\chapno.\the\figureno?}}\fi\fi
\global\expandafter\edef\csname FIGLABEL#1\endcsname
{\the\chapno.\the\figureno?}\iffigurechapternumbers\chapfolio.\fi
\ifproofmode\llap{\hbox{\marginstyle#1
\kern1.2truein}}\relax\fi\the\figureno}

\def\figref#1{\hbox{\ifundefined{FIGLABEL#1}!!!!\ifproofmode\ifforwardreference%
\else\write16{ ***Undefined Figure Reference #1*** }\fi
\else\write16{ ***Undefined Figure Reference #1*** }\fi
\else\edef\LABxx{\getlabel{FIGLABEL#1}}%
\def\LAByy{\expandafter\stripchap\LABxx}\iffigurechapternumbers%
\chapshow{\LAByy}.\expandafter\stripeq\LABxx%
\else\ifnum \number\LAByy=\chapno\relax\expandafter\stripeq\LABxx%
\else\chapshow{\LAByy}.\expandafter\stripeq\LABxx\fi\fi\fi}%
\ifproofmode\write2{Figure #1}\fi}

\def\tabnum{\global\advance\tableno by 1
\relax\iftablechapternumbers\chapfolio.\fi\the\tableno}

\def\tablabel#1{\global\advance\tableno by 1
\relax\ifproofmode\ifforwardreference
\immediate\write1{\noexpand\expandafter\noexpand\def
\noexpand\csname TABLABEL#1\endcsname{\the\chapno.\the\tableno?}}\fi\fi
\global\expandafter\edef\csname TABLABEL#1\endcsname
{\the\chapno.\the\tableno?}\iftablechapternumbers\chapfolio.\fi
\ifproofmode\llap{\hbox{\marginstyle#1
\kern1.2truein}}\relax\fi\the\tableno}

\def\tabref#1{\hbox{\ifundefined{TABLABEL#1}!!!!\ifproofmode\ifforwardreference%
\else\write16{ ***Undefined Table Reference #1*** }\fi
\else\write16{ ***Undefined Table Reference #1*** }\fi
\else\edef\LABtt{\getlabel{TABLABEL#1}}%
\def\LABTT{\expandafter\stripchap\LABtt}\iftablechapternumbers%
\chapshow{\LABTT}.\expandafter\stripeq\LABtt%
\else\ifnum\number\LABTT=\chapno\relax\expandafter\stripeq\LABtt%
\else\chapshow{\LABTT}.\expandafter\stripeq\LABtt\fi\fi\fi}%
\ifproofmode\write2{Table#1}\fi}

\newdimen\sectionskip     \sectionskip=20truept
\newcount\sectno
\def\section#1#2{\sectno=0 \null\vskip\sectionskip
    \centerline{\chaplabel{#1}.~~{\bf#2}}\nobreak\vskip.2truein
    \noindent\ignorespaces}

\def\advancesectno{\global\advance\sectno by 1}
\def\sectfolio{\number\sectno}
\def\subsection#1{\goodbreak\advancesectno\null\vskip10pt
                  \noindent\chapfolio.~\sectfolio.~{\bf #1}
                  \nobreak\vskip.05truein\noindent\ignorespaces}

\def\uttg#1{\null\vskip.1truein
    \ifproofmode \line{\hfill{\bf Draft}:
    UTTG--{#1}--\number\year}\line{\hfill\today}
    \else \line{\hfill UTTG--{#1}--\number\year}
    \line{\hfill\ifcase\month\or January\or February\or March\or April\or
May\or June
    \or July\or August\or September\or October\or November\or December\fi
    \space\number\year}\fi}

\def\getlabel#1{\csname#1\endcsname}
\def\ifundefined#1{\expandafter\ifx\csname#1\endcsname\relax}
\def\stripchap#1.#2?{#1}
\def\stripeq#1.#2?{#2}

%
\catcode`@=11 
\def\space@ver#1{\let\@sf=\empty\ifmmode#1\else\ifhmode%
\edef\@sf{\spacefactor=\the\spacefactor}\unskip${}#1$\relax\fi\fi}
\newcount\referencecount     \referencecount=0
\newif\ifreferenceopen       \newwrite\referencewrite
\newtoks\rw@toks
\def\refmark#1{\relax[#1]}
\def\refend{\refmark{\number\referencecount}}
\newcount\lastrefsbegincount \lastrefsbegincount=0
\def\refsend{\refmark{\count255=\referencecount%
\advance\count255 by -\lastrefsbegincount%
\ifcase\count255 \number\referencecount%
\or\number\lastrefsbegincount,\number\referencecount%
\else\number\lastrefsbegincount-\number\referencecount\fi}}
\def\refch@ck{\chardef\rw@write=\referencewrite
\ifreferenceopen\else\referenceopentrue
\immediate\openout\referencewrite=referenc.texauxil \fi}
%
{\catcode`\^^M=\active 
  \gdef\obeyendofline{\catcode`\^^M\active \let^^M\ }}%
%
{\catcode`\^^M=\active 
  \gdef\ignoreendofline{\catcode`\^^M=5}}
{\obeyendofline\gdef\rw@start#1{\def\t@st{#1}\ifx\t@st\blankend%
\endgroup\@sf\relax\else\ifx\t@st\bl@nkend\endgroup\@sf\relax%
\else\rw@begin#1
\backtotext
\fi\fi}}
{\obeyendofline\gdef\rw@begin#1
{\def\n@xt{#1}\rw@toks={#1}\relax%
\rw@next}}
\def\blankend{}
{\obeylines\gdef\bl@nkend{
}}
\newif\iffirstrefline  \firstreflinetrue
\def\rwr@teswitch{\ifx\n@xt\blankend\let\n@xt=\rw@begin%
\else\iffirstrefline\global\firstreflinefalse%
\immediate\write\rw@write{\noexpand\obeyendofline\the\rw@toks}%
\let\n@xt=\rw@begin%
\else\ifx\n@xt\rw@@d \def\n@xt{\immediate\write\rw@write{%
\noexpand\ignoreendofline}\endgroup\@sf}%
\else\immediate\write\rw@write{\the\rw@toks}%
\let\n@xt=\rw@begin\fi\fi\fi}
\def\rw@next{\rwr@teswitch\n@xt}
\def\rw@@d{\backtotext} \let\rw@end=\relax
\let\backtotext=\relax

\newdimen\refindent     \refindent=30pt
\def\Textindent#1{\noindent\llap{#1\enspace}\ignorespaces}
\def\refitem#1{\par\hangafter=0 \hangindent=\refindent\Textindent{#1}}
\def\REFNUM#1{\space@ver{}\refch@ck\firstreflinetrue%
\global\advance\referencecount by 1 \xdef#1{\the\referencecount}}
\def\refnum#1{\space@ver{}\refch@ck\firstreflinetrue%
\global\advance\referencecount by 1\xdef#1{\the\referencecount}\refend}

\def\REF#1{\REFNUM#1%
\immediate\write\referencewrite{%
\noexpand\refitem{#1.}}%
\begingroup\obeyendofline\rw@start}
\def\ref{\refnum\?%
\immediate\write\referencewrite{\noexpand\refitem{\?.}}%
\begingroup\obeyendofline\rw@start}
\def\Ref#1{\refnum#1%
\immediate\write\referencewrite{\noexpand\refitem{#1.}}%
\begingroup\obeyendofline\rw@start}
\def\REFS#1{\REFNUM#1\global\lastrefsbegincount=\referencecount%
\immediate\write\referencewrite{\noexpand\refitem{#1.}}%
\begingroup\obeyendofline\rw@start}

\def\REFSCON#1{\REF#1}

\def\cite#1{\refmark#1}
\catcode`@=12 
%
\input epsf.tex
\baselineskip=15pt plus 1pt minus 1pt
\parskip=5pt
\chapternumberstrue
\figurechapternumberstrue
\tablechapternumberstrue
\ifproofmode
\immediate\openout2=allcrossreferfile \fi
\ifforwardreference\input labelfile
\ifproofmode\immediate\openout1=labelfile \fi\fi
\hfuzz=2pt
\vfuzz=2pt


\def\hourandminute{\count255=\time\divide\count255 by 60
\xdef\hour{\number\count255}
\multiply\count255 by -60\advance\count255 by\time
\hour:\ifnum\count255<10 0\fi\the\count255}

\def\subsection#1{\goodbreak\advancesectno\null\vskip10pt
                  \noindent{\it \chapfolio.\sectfolio.~#1}
                  \nobreak\vskip.05truein\noindent\ignorespaces}
\def\cite#1{\refmark{#1}}
\def\\{\hfill\break}

\def\point#1{\noindent\setbox0=\hbox{#1}\kern-\wd0\box0}

\nopagenumbers\pageno=0
\rightline{\eightrm }\vskip-5pt
\rightline{\eightrm hep-th/9812253}\vskip-5pt
\rightline{\eightrm December 31, 1998}

\vskip1.2truein
\centerline{\seventeenrm  On the Hypermultiplet Moduli Space}
\vskip10pt
\centerline{\seventeenrm of Heterotic Compactifications with Small Instantons}
\vskip55pt
\centerline{\csc Eugene~Perevalov$^1$}
\vfootnote{$^{\eightrm 1}$}{\eightrm pereval@math.harvard.edu}

\vskip.8truein\bigskip
\centerline{\it Department of Mathematics}
\centerline{\it Harvard University}
\centerline{\it Camdridge, MA 02138, USA}
\vskip.6in\bigskip
\nobreak\vbox{
\centerline{\bf ABSTRACT}
\vskip.25truein
\noindent{We explore a relation between four-dimensional $N=2$ heterotic vacua
induced by Mirror Symmetry via Heterotic/Type II duality. It allows us to
compute the $\alpha'$ corrections to the hypermultiplet moduli space of 
heterotic compactifications on $K3\times T^2$ in the limit of large base of the
elliptic $K3$. We concentrate on the case of point-like instantons on
orbifold singularities leading to low-dimensional hypermultiplet moduli spaces.}
}
\newpage
\pageno=1
\headline={\ifproofmode\hfil\eightrm draft:\ \today\
\hourandminute\else\hfil\fi}
\footline={\rm\hfil\folio\hfil}
\section{intro}{Introduction}
When a heterotic string is compactified on a $K3\times T^2$ manifold the result
is an $N=2$ supergravity coupled to Yang-Mills with matter. The moduli space of
the above theory is parametrized by the VEVs of scalar components of vector and
hypermultiplets and locally has the following form.
$${\cal M}\cong {\cal M}_H\times {\cal M}_V,$$
where ${\cal M}_H$ is a quaternionic manifold and ${\cal M}_V$ is a special \K\
manifold. 
The geometry of the moduli space encodes the information about the effective
low-energy theory and is of primary interest. In general, it receives corrections
of two types: due to finite size of the compactification manifold and to nonzero
string coupling ($\alpha'$ and $g_s$ corrections, respectively). However, since
the string coupling is determined by the VEV of the dilaton which is a scalar
component of a vector multiplet
in heterotic compactifications, ${\cal M}_H$ is unaffected by $g_s$ corrections.
${\cal M}_V$ on the other hand is expected to be corrected by the
quantum effects. What allows one to compute these corrections and obtain exact
expressions for the geometry of ${\cal M}_V$ are two kinds of dualities believed
to hold in string theory: Mirror Symmetry (see \eg \Ref\Y{S.-T. Yau, editor,
Essays on Mirror Manifolds, International Press, 1992}\
and Heterotic/Type II duality
\REFS\KV{S.~Kachru and C.~Vafa, \npb{450} (1995) (69), hep-th/9505105.}
\REFSCON\FNSV{S. Ferrara, J. A. Harvey, A. Strominger and C. Vafa,
\plb{361} (1995) 59,\\ hep-th/9505162.}
\refsend .

Heterotic/Type II duality relates heterotic strings compactified on
$K3\times T^2$
to Type IIA strings on a \cyt. The key fact which allows us to sum up $g_s$
corrections to ${\cal M}_V$ is that in type IIA compactifications the dilaton
resides in a hypermultiplet and (in the low energy limit) cannot affect the
geometry of vector multiplets. Thus if one can find a type IIA compactification
dual to a heterotic model in question the corresponding ${\ca M}_V$ will be
$g_s$-exact on the type IIA side. At this stage however one still has to
compute the $\a'$ corrections since for type IIA vector multiplets correspond
to \K\ class parameters of the \cyt\ and those are corrected by the world-sheet
instantons. The predicament is resolved by turning to mirror symmetry
conjecture which in its "full" form states that the type IIA string
compactified on a \cym\ $M$ is physically equivalent to the type IIB string
compactified on the mirror manifold $W$. Thus the \K\ class parameters of $M$
get replaced by the complex structure parameters of $W$ which do not suffer from
the corrections due to finite size of the manifold. In this way one can manage
to obtain the exact low energy form of the vector multiplet moduli space
of a $N=2$ heterotic model in four dimensions.

The question we address in this paper is given such a heterotic
compactification how to compute $\a'$ corrections to the classical moduli
space of hypermultiplets. For this purpose we are going to use the concept
of a {\bf c}-map introduced
in~\Ref\CFG{S. Cecotti, S. Ferrara and L. Girardello, Intern. Journ.
Mod. Phys. {\bf A4} 10 (1989) 2475.}. The {\bf c}-map relates the vector
multiplet moduli space resulting from a compactfication of type IIA theory
on a \cyt\ $M$ to the hypermultiplet moduli space obtained by the
compactification of type IIB theory on the same manifold (or, equivalently,
by type IIA on the mirror $W$). Thus it converts special \K\ manifolds into
quaternionic ones. For our purpose we need a heterotic version of the
{\bf c}-map. Namely given a heterotic model in four dimensions we would like
to know what the "mirror" model is (the original one being equivalent to the
type IIA on a \cyt\ $M$ the "mirror" one is defined as the dual to the type
IIA string theory compactified on the mirror manifold $W$). The answer to this
question is provided
in~\Ref\PR{E. Perevalov and G. Rajesh, Phys. Rev. Lett., {\bf 79} (1997)
2931, hep-th/9706005.}\ (see also~\Ref\BM{P. Berglund and P. Mayr,
hep-th/9811217.})
and is roughly speaking that in order to obtain the "mirror"
model we have to replace all
finite size instantons with structure group $G$ by point-like instantons
\REFS\Wsi{E. Witten, \npb{460} (1996) 541, hep-th/9511030.}
\REFSCON\AM{P.~S.~Aspinwall and D.~R.~Morrison, \npb{503} (1997)
533, hep-th/9705104.}\refsend\
sitting
on an orbifold singularity of the $K3$ corresponding to $G$ and
vice versa.\Footnote{There as well as in this paper we restrict ourselves
to the heterotic compactifications which can be lifted to six dimensions, \ie
with trivial vector bundles over the $T^2$ component of the compactification
manifold. More precisely, we consider models where the bundle over
$K3\times T^2$ has the form of a direct product of bundles over the
corresponding factors of the compactification space, with the factor over
the $T^2$ being trivial (no Wilson lines).}
Thus we can using the {\bf c}-map compute $\a'$ corrections to the metric on the
moduli space of hypermultiplets of our original model from that on the moduli
space of vector multiplets of the "mirror" model. An important point to note
here is that by doing this we do not obtain the {\it exact} metric on the
moduli space since the {\bf c}-map does not capture the quantum effects on the
type II side which, under the type II/heterotic duality, become part of
$\a'$ corrections in the heterotic model.\Footnote{I am grateful to
E. Witten for pointing this out to me.} Recalling that the type IIA dilaton
is mapped to the size of the base of the elliptic fibration $S_H\rightarrow B_H$
where $S_H$ is the heterotic $K3$
\Ref\Ahyp{P. S. Aspinwall, hep-th/9802194.},
we conclude that this method allows us to compute the $\a'$ corrections 
when the base $B_H$ is large with additional corrections arising away from
this limit.

This paper is organized as follows. In \SS{2} we briefly review how the exact
expressions on the moduli space of vector multiplets can be obtained with
the help of Heterotic/TypeII duality and Mirror Symmetry. In \SS{3} we describe
the {\bf c}-map. In \SS{4} we use the {\bf c}-map and the results of  
\cite{\PR} to compute the $\a'$ corrections to the metric (or rather the
holomorphic function which encodes it) on the moduli space of hypermultiplets
of heterotic compactifications in the regime of large $B_H$.
In particular we concentrate on the simpler
case of point-like instantons where the classical moduli space is just that
of sigma models on $K3$ orbifolds and we do not have to deal with
the bundle moduli. \SS{5} summarizes our results.

\section{vect}{Exact results on vector multiplet moduli space}
In this section we briefly summarize the use of string dualities for the purpose
of obtaining exact expressions for the moduli space of vector multiplets
in four-dimensional heterotic compactifications with $N=2$.
As we have already mentioned ${\ca M}_V$ has the structure of a special \K\
manifold meaning that both the \K\ potential and the Wilsonian gauge couplings
of the vector multiplets in the $N=2$ effective lagrangian are determined by a
single holomorphic function of the moduli --- the prepotential $F(X)$ which
is (in supergravity) homogeneous of second degree in complex fields $X^I$
where $I=0,1,\ldots n_V$ ($n_V$ being the number of vector multiplets).
The \K\ potential is given in terms of the prepotential by
$$K=-\log\left(i{\overline X}^IF_I(X)-iX^I{\overline F}_I
({\overline X}^I)\right),\eqlabel{Kpotx}$$
where $F_I(X)\equiv {\partial F(X)\over \partial X^I}$. In fact the physical
moduli parametrize an $n_V$-dimensional manifold. A convenient choice of
coordinates is $X^0(z)=1$, $X^A(z)=z^A$, $A=1,\ldots n_V$. In these (special)
coordinates the \K\ potential becomes
$$K=-\log\left(2({\ca F}+{\overline {\ca F}})-(z^A-{\overline z}^A)({\ca F}_A
-{\overline {\ca F}}_A)\right),\eqlabel{Kpotz}$$
where now {\ca F}$(z)$ is an arbitrary holomorphic function of $z^A$ related
to $F(X)$ via $F(X)=i(X^0)^2{\ca F}(z)$.

For $N=2$ heterotic vacua the prepotential  receives contributions at the
string tree, one loop and the non-pertubatitive level:
$${\ca F}^{(0)}(S,M)+{\ca F}^{(1)}(M)+{\ca F}^{({\rm np})}(e^{-8\p^2S},M),
\eqlabel{hetf}$$
where $S$ stands for the dilaton and $M$ for the rest of the vector moduli.
The fact that the dilaton arises in the univesal sector together with the
Peccei-Quinn symmetry and the form of the \K\ potential~\eqref{Kpotz} constrain
the tree level contribution to the prepotential
to be~\Ref\FvP{S. Ferrara and A. van Proeyen, Class. Quant. Grav. {\bf 6} (1989)
243.} 
$${\ca F}=-S(TU-\phi^i\phi^i),\eqlabel{ftree},$$
where $T$ and $U$ are the usual torus moduli and $i=1,\ldots n_V-3$.
It follows that the the tree level \K\ potential has the form
$$K^{(0)}=-\log(S+{\overline S})-\log({\rm Re} T{\rm Re} U-
{\rm Re}\phi^i{\rm Re}\phi^i)\eqlabel{Ktree},$$
and the metric derived from it is the natural metric on the coset space
$${\ca M}_V^{(0)}={SU(1,1)\over U(1)}\times {SO(2,n_V-1)\over
SO(2)\times SO(n_V-1)},\eqlabel{treemetric}$$
where the first factor corresponds to the dilaton.
The one loop contribution to the prepotential can be computed using the
perturbative quantum symmetries and the singularity structure corresponding
to the gauge symmetry enhancement loci
(see \eg~\Ref\L{D. L\"{u}st, hep-th/9803072 and references therein.}). But we
will use the dual type IIA model to obtain the fully corrected form of the
prepotential. In order to do that we need to find the type II compactification
dual to the heterotic one we are considering. Having done that  we can write the
corresponding exact form of the prepotential as ($t_A=-iz^A$ are the \K\ class
parameters and $q_A=e^{-2\p t_A}$).
$${\ca F}=-{1\over 6}C_{ABC}t_At_Bt_C+{1\over (2\p)^3}
\sum_{d_1,\ldots d_n}n_{d_1,\ldots d_n}
{\rm Li}_3(\prod_{i=1}^nq_A^{d_A}).\eqlabel{fdual}$$
Here the $C_{ABC}$ are the classical intersection numbers of the \cyt\ and  
$n_{d_1,\ldots d_n}$ are the rational instanton numbers of multidegree
$d_1,\ldots d_n$.  Note that the cubic part of the prepotential generated at
the tree level on the type II side corresponds to the sum of the tree and one
loop level contributions on the heterotic side. The rational instanton numbers
can be found by replacing, by mirror symmetry, the type IIA compactified on $M$
by type IIB on the mirror \cyt\ $W$ where the vector mutiplets correspond to
the complex structure parameters and the moduli space is not corrected by
the world-sheet instantons and is exact at the string tree level. They have
been calculated explicitly in a number of low-dimensional examples
\REFS\COGP{P. Candelas, X. de la Ossa, P. Green and L. Parkes, \npb359
(1991) 21.}
\REFSCON\COFKM{P. Candelas, X. de la Ossa, A. Font, S. Katz and D. Morrison,
\npb416 (1994) 481.}
\REFSCON\HKTY{S. Hosono, A. Klemm, S. Theisen and S.-T. Yau,
\cmp167 (1995) 301.}
\refsend.

Consider the heterotic vacuum obtained by taking a vector bundle with structure
group $E_8\times E_8$ with the instanton number 14 in one factor and 10 in
the other. Then the
primordial gauge symmetry is broken completly and we get a model with 3 vector
multiplets (including the dilaton) in four dimensions (the so-called $STU$
model). The number of neutral hypermultiplets can be computed from the
formula
$$n_H=h_1k_1-{\rm dim}G_1+h_2k_2-{\rm dim}G_2+20\eqlabel{nH}$$
with $h_i$, $k_i$, $G_i$ being the dual Coxeter number, the instanton number
and the structure group for the two components of the vector bundle, and the
universal 20 arising from the number of the (smooth) $K3$ moduli. In our case
this gives 244.
The dual type II vacuum results from the compactificaton of type IIA
string theory on the \cyt\ which can be constructed as a hypersurface in the
toric variety defined by the data displayed in
Table~\tabref{weights}\Footnote{One starts
with homogeneous coordinates $s,t,u,v,x,y,w$ removes the loci $\{s=t=0\}$,
$\{u=v=0\}$, $\{x=y=w=0\}$, and takes the quotent by three scalings
$(\l, \m, \n)$ with the exponents from Table~\tabref{weights}.} for $n=2$.
The resulting \cym\ is an elliptic fibration over the Hirzebruch surface
$\IF_2$.
\midinsert
$$\def\skip{\hskip10pt}
\def\extraspace{\omit{\vrule height5pt}&&&&&&&&&\cr}
\vbox{\offinterlineskip\halign{\strut # height 12pt depth 6pt
&\hfil\skip \eightrm $#$\skip\hfil
&\hfil\skip \eightrm $#$\skip\hfil
&\hfil\skip \eightrm $#$\skip\hfil 
&\hfil\skip \eightrm $#$\skip\hfil
&\hfil\skip \eightrm $#$\skip\hfil
&\hfil\skip \eightrm $#$\skip\hfil
&\hfil\skip \eightrm $#$\skip\hfil
&\hfil\skip \eightrm $#$\skip\hfil
&\hfil\skip \eightrm $#$\skip\hfil\vrule\cr
\noalign{\hrule}
\omit{\vrule height3pt}&&&&&&&&&\cr
\vrule&&s&t&u&v&x&y&w&$Degrees$\cr
\omit{\vrule height3pt}&&&&&&&&&\cr
\noalign{\hrule}
\omit{\vrule height5pt}&&&&&&&&&\cr
\vrule&\l&1&1&n&0&2n+4&3n+6&0&6n+12\cr
\extraspace
\vrule&\m&0&0&1&1&4&6&0&12\cr
\extraspace
\vrule&\n&0&0&0&0&2&3&1&6\cr
\omit{\vrule height5pt}&&&&&&&&&\cr
\noalign{\hrule}
}}
$$
\nobreak\tablecaption{weights}{The scaling weights for the toric variety.}
\bigskip
\endinsert
The Hodge numbers of the
\cyt\ are $h_{11}=3$, $h_{21}=243$, in agreement with the heterotic spectrum.
The classical intersection numbers for this \cym\ are easily calculated and
the resulting cubic prepotential is
$$-{\ca F}_{\rm cubic}={4\over 3}t_1^3+t_1^2t_2+2t_1^2t_3+t_1t_2t_3
+t_1t_3^2.\eqlabel{fcubic}$$
The rational instanton numbers for this model have been
calculated in~\cite{\HKTY} and in~\Ref\BKKM{P. Berglund, S.Katz, A. Klemm
and P. Mayr, \npb{483} (1997) 209, hep-th/9605154.}.

\section{cmap}{Review of the {\bf c}-map}
Originally the {\bf c}-map was defined in~\cite{\CFG} as a one which transforms
the low-energy effective lagrangian for a type IIA superstring into the
corresponding lagrangian for a type IIB superstring compactified on the same
(2,2) superconformal system. If we restrict ourselves to cases where the
abstract superconformal system can be represented by a \cym\ with Hodge numbers
$h_{11}$ and $h_{21}$ then the spectrum of a type IIA compactification contains
$h_{11}$ vector multiplets and $h_{21}+1$ hypermultiplets (1 counting the
dilaton) and the corresponding numbers for the type IIB compactifications are
$h_{21}$ and $h_{11}+1$, respectively.
Since in $N=2$ supergravity the scalars of the
vector multiplets parametrize a special \K\ manifold (a \K\ manifold with a
holomorphic prepotential) and the scalars in the hypermultiplets a quaternionic
manifold, the {\bf c}-map can be seen as a transformation sending a special \K\
manifold of complex dimension $h_{11}$ into a quaternionic manifold of real
dimension $4(h_{11}+1)$ and a quaternionic manifold of dimension
$4(h_{21}+1)$ into a special \K\ manifold of dimension $h_{21}$, \ie the
{\bf c}-map has the general form
$${\bf c}: K_{h_{11}}\otimes Q_{h_{21}+1}\rightarrow K_{h_{21}}
\otimes Q_{h_{11}+1},\eqlabel{cmapgen}$$
where $K_n$ is the set of all special \K\ $n$-folds and $Q_m$ the set of
quaternionic spaces of quaternionic dimension $m$ (real dimension $4m$).
But the fundamental operation as was shown in~\cite{\CFG} is not {\bf c}
but ${\bf s}_n$  
$${\bf s}_n: K_n \rightarrow Q_{n+1}\eqlabel{snmapgen}$$
mapping a special \K\ $n$-fold into a quaternionic space of real dimension
$4(n+1)$. The image of the map ${\bf s}_n$ is some proper subset of 
$Q_{n+1}$ denoted by ${\tilde Q}_{n+1}$ and called dual-quaternionic:
$${\tilde Q}_{n+1}\equiv {\rm Im}({\bf s}_n)\subset Q_{n+1}.\eqlabel{duquat}$$
Thus ${\bf s}_n$ is an isomorphism between $K_n$ and ${\tilde Q}_{n+1}$. It was
proven
in~\cite{\CFG} that in type II theories the hypermultiplet moduli spaces belong
to ${\tilde Q}_n$ for appropriate $n$. The low-energy effective lagrangian then
is specified by a point in the following spaces:
$$K_{h_{11}}\otimes {\tilde Q}_{h_{21}+1}\hskip20pt ({\rm type\ IIA})$$
$${\tilde Q}_{h_{11}+1}\otimes K_{h_{21}}\hskip20pt ({\rm type\ IIB})
\eqlabel{efL}$$

For any special \K\ space the {\bf c}-map can be constructed in the following
way. Given a four-dimensional $N=2$ supergravity reduce it dimensionally to
three dimensions. The result is an $N=4$ supergravity. After this dimensional
reducton, the hypermultiplet moduli in three dimensions parametrize the same
quaternionic manifold they did in four dimensions. As for vector moduli, we
get one extra scalar from the fourth component of the vector. And moreover,
in three dimensions, a vector is dual to a scalar so we get one more. Thus
after the duality transformation the bosonic part of of the vector multiplet
sector reduces to a theory of $4n+4$ scalars if we have started with $n$ vector
multiplets in four dimensions. The four additional scalars come from the
gravitational sector. These $4n+4$ scalars parametrize a quaternionic manifold
as required by $N=4$ supersymmetry. Thus the resulting moduli space in three
dimensions has the form of a product of two quaternionic manifolds:
$${\ca M}_3\cong {\ca M}_3^{(1)}\times {\ca M}_3^{(2)},\eqlabel{modsred}$$
where ${\ca M}_3^{(1)}$ is the 4D hypermultiplet manifold, ${\ca M}_3^{(2)}$
arises from the 4D vector multiplets. Then the {\bf c}-map just corresponds to
the interchange of ${\ca M}_3^{(1)}$ and ${\ca M}_3^{(2)}$ or, more precisely,
the result of the action of ${\bf s}_n$ on the special \K\ space in question
is ${\ca M}_3^{(2)}$, the quaternionic space resulting from the dimensional
reduction of the vector multiplet sector to 3D.

Before we give the explicit result of this transformation in a general case
let us consider the simpler one of symmetric special \K\ spaces. All such spaces
allowed in $N=2$ supergravity~\Ref\CvP{E. Cremmer and A. van Proeyen, Class.
Quant. Grav. {\bf 2} (1985) 445.}\ are listed in the first column of
Table~\tabref{cmapsym}. There are two general families and five exceptional
models related to the Jordan algebras. In these cases the special \K\ and
quaternionic spaces related by the {\bf c}-map are simply those associated
to the same Jordan algebra. This fact was noticed by physicists
in~\Ref\GST{M. Gunyadin, G. Sierra and P. K. Townsend, \plb{133} (1983) 72.}.
For example the \K\ space ${U(3,3)\over U(3)\times U(3)}$ and the quaternionic
one ${E_{6(+2)}\over SU(2)\times SU(6)}$ are associated to the Jordan algebra
$J^C_{\hskip4pt 3}$, the Jordan algebra of the hermitian $3\times 3$ complex
matrices.
\midinsert
$$\def\skip{\hskip25pt}
\def\extraspace{\omit{\vrule height10pt}&&&\cr}
\vbox{\offinterlineskip\halign{\strut # height 12pt depth 6pt
&\hfil\skip \eightrm $#$\skip\hfil\vrule
&\hfil\skip \eightrm $#$\skip\hfil\vrule
&\hfil\skip \eightrm $#$\skip\hfil\vrule \cr
\noalign{\hrule}
\omit{\vrule height3pt}&&&\cr
\vrule&$\K\ space,$\hskip5pt{\ca M}&{\rm dim}_{\IC}{\ca M}&
$Quaternionic space$\cr
\omit{\vrule height3pt}&&&\cr
\noalign{\hrule\vskip3pt\hrule}
\omit{\vrule height5pt}&&&\cr
\vrule&{U(1,n)\over U(n)\times U(1)}&n&{U(2,n+1)\over U(2)\times U(n+1)}\cr
\extraspace
\vrule&{SU(1,1)\over U(1)}\times {SO(n-1,2)\over SO(n-1)\times SO(2)}&n\ge 2&
{SO(n+1,4)\over SO(n+1)\times SO(4)}\cr
\extraspace
\vrule&{SU(1,1)\over U(1)}&1&{G_{2(+2)}\over SO(4)}\cr
\extraspace
\vrule&{Sp(6,\IR)\over U(3)}&6&{F_{4(+4)}\over USp(6)\times SU(2)}\cr
\extraspace
\vrule&{U(3,3)\over U(3)\times U(3)}&9&{E_{6(+2)}\over SU(6)\times SU(2)}\cr
\extraspace
\vrule&{SO^{\ast}(12)\over U(6)}&15&{E_{7(-5)}\over SO(12)\times SU(2)}\cr
\extraspace
\vrule&{E_{7(-26)}\over E_6\times SO(2)}&27&{E_{8(-24)}\over E_7\times SU(2)}\cr
\omit{\vrule height5pt}&&&\cr
\noalign{\hrule}
}}
$$
\nobreak\tablecaption{cmapsym}{The {\bf c}-map for symmetric \K\ spaces.}
\bigskip
\endinsert

A full classification of the quaternionic spaces with a transitive solvable
group of motions (normal quaternionic spaces)
was given in mathematics literature~\Ref\A{D. V. Alekseevskii,
Math USSR Izvestija {\bf 9} (1975) 297.}.
More precisely, the normal dual-quaternionic spaces were characterized and
their images under the inverse {\bf s}-map constructed. This approach is more
general than that based on Jordan algebras since it provides the {\bf c}-map
for homogeneous and not just symmetric spaces.

The {\bf c}-map in the general case was worked out in~\Ref\FS{S. Ferrara and
S. Sabharwal, \npb{332} (1990) 317.}. Let us recall the results.

The bosonic part of of the $N=2$ lagrangian for vector multiplets is
\REFS\dWvP{B. de Wit and A. van Proeyen, \npb{245} (1984) 89.}
\REFSCON\Cetal{E. Cremmer et. al. \npb{250} (1985) 385.}\refsend 
$${\ca L}={1\over 2}R-K_{A{\overline B}}\partial_{\m}
z^A\partial^{\m}{\overline z}^B+
{1\over 4}{\rm Re}{\ca N}_{IJ}F^I_{\m\n}F^{\m\n J}+
{1\over 4}{\rm Im}{\ca N}_{IJ}F_{\m\n}^I{\tilde F}^{\m\n J},\eqlabel{vmulL}$$
with
$${\ca N}_{IJ}={1\over 2}{\overline F}_{IJ}-{(Nz)_I(Nz)_J\over (zNz)},
\hskip20pt N_{IJ}={1\over 2}(F_{IJ}+{\overline F}_{IJ})$$
and $K$ as in~\eqref{Kpotz}.

The dimensional reduction to three dimensions yields the lagrangian for the
scalar manifold of the following form
$$\eqalign{{\ca L}=&-K_{A{\overline B}}\partial_{\m}z^A\partial^{\m}
{\overline z}^B-
\bigl[S+{\overline S}+{1\over 2}(C+{\overline C})R^{-1}
(C+{\overline C})\bigr]^{-2}\cr
&\times\big|\partial_{\m}S+(C+{\overline C})R^{-1}\partial_{\m}C
-{1\over 4}(C+{\overline C})R^{-1}\partial_{\m}{\ca N}R^{-1}(C+{\overline C})
\big|^2\cr
&+\bigl[S+{\overline S}+{1\over 2}(C+{\overline C})R^{-1}
(C+{\overline C})\bigr]^{-1}
\bigl(\partial_{\m}C-{1\over 2}\partial_{\m}{\ca N}R^{-1}
(C+{\overline C})\bigr)\cr
&\times R^{-1}\bigl(\partial_{\m}{\overline C}-{1\over 2}\partial_{\m}
{\overline {\ca N}}R^{-1}(C+{\overline C})\bigr),\cr}\eqlabel{dredL}$$
where $R_{IJ}={\rm Re}\ca N_{IJ}$.

Eq.~\eqref{dredL} defines a manifold for the $2(n+1)$ complex scalar fields
$S$, $z^A$ and $C_I$ which is a dual-quaternionic manifold ${\tilde \ca M_V}$ of
(real) dimension $4n_V+4$ which is the image under the ${\bf s}_n$ map of the
special \K\ manifold ${\ca M}_V$ of complex dimension $n_V$. \eqref{dredL}
can be rewritten as
$$\eqalign{{\ca L}=&-K_{A{\overline B}}\partial_{\m}z^A\partial^{\m}
{\overline z}^B-\tilde K_{S\overline S}D_{\m}SD^{\m}\overline S
-\tilde K_{S\overline C_I}D_{\m}SD^{\m}\overline C_I-\tilde K_{C_I\overline S}
D_{\m}C_ID^{\m}\overline S\cr
&-\tilde K_{C_I\overline C_J}D_{\m}C_ID^{\m}\overline C_J}\eqlabel{dredLm},$$
where
$$D_{\m}C=\partial_{\m}C-{1\over 2}\partial_{\m}{\ca N}R^{-1}(C+\overline C),$$
$$D_{\m}S=\partial_{\m}S+{1\over 4}(C+\overline C)R^{-1}\partial_{\m}{\ca N}
R^{-1}(C+\overline C),$$
$$\tilde K=-\log\bigl(S+\overline S+{1\over 2}(C+\overline C)R^{-1}
(C+\overline C)\bigr).$$

The resulting space enjoys the following properties:
\item{$\bullet$}At each fixed $z^A$ the Ramond-Ramond scalars $C_I$ parametrize
a ${SU(1,n_V+2)\over S(U(1)\times U(n_V+2))}$ manifold.
\item{$\bullet$}At each point (${\rm Re}C_I=0$, ${\rm Im}C_I={\rm const}$) the
$z^A$, $S$ fields parametrize a ${SU(1,1)\over U(1)}\times \ca M_V$ manifold
where \ca M$_V$ is the original \K\ manifold of the vector multiplets.
\item{$\bullet$}The quaternionic manifold ${\tilde \ca M}_V$ has at least
$2n_V+4$ isometries acting on all but $z^A$ coordinates.

\section{het}{Moduli space of hypermultiplets in heterotic compactifications}
Our aim in this section is twofold: to give some evidence in favor of the
conjecture of~\cite{\PR} and to use it in conjunction with the exact results
on the moduli space of vector multiplets and the {\bf c}-map transformation
to compute the $\a'$ corrections to the moduli space of hypermultiplets
in the regime of large $B_H$. For the
sake of simplicity, we will discuss only the models where all instantons are
point-like although our method should work as well for any heterotic $K3\times
T^2$ compactification.\Footnote{Provided, as we have already mentioned in \SS{1},
that the bundle over $K3\times T^2$ factorizes into a product of bundles over
$K3$ and $T^2$, the latter being trivial.}

We will illustrate our approach by one particular example. Consider a heterotic
obtained by taking a $K3$ with two $E_8$ singularities with 14 point-like
instantons located at one singularity and 10 at the other.
The resulting spectrum in four
dimensions consists of 243 vector multiplets and 4 hypermultiplets.\Footnote{The
corresponding vacuum in six dimensions obtained by decompactifying the $T^2$
has 97 tensor multiplets, $E_8^8\times F_4^8\times G_2^{16}\times SU(2)^{16}$
gauge group (for a total of 2672 vector multiplets) with the hypermulplet
content including~
\Ref\CPRm{P. Candelas, E. Perevalov and G. Rajesh, \npb{519} (1998) 225.}
$\lbrace {1\over 2}(\bf 7,\bf 2)+{1\over 2}(\bf 1, \bf 2)\rbrace $ for each
of the 16 factors of $G_2\times SU(2)$ as well as 4 neutral hypermultiplets.
Thus for this vacuum, $H=H_c+H_0=128+4=132$ and the anomaly cancellation
condition is satisfied as $H-V=132-2672=-2540=273-29\cdot 97=273-29T$.}
\subsection{The classical moduli space}
The classical hypermultiplet moduli space in this example is simply that of
sigma models on a $K3$ orbifold with two $E_8$ singularities which can be
determined as follows. As is well known (see \eg~\Ref\Arev{P. S. Aspinwall,
hep-th/9611137.}) $H_2(S,\IZ)$ for a $K3$ surface $S$ is a 22-dimensional
self-dual even lattice of signature (3,19) and
one can choose a basis of 2-cycles so that the inner product on the basis
elements of $H_2(S,\IZ)$ forms the matrix
$$\def\skip{\hskip5pt}
\left(\matrix{\left(\matrix{&&\cr &-E_8&\cr &&\cr}\right)&&&&\cr
               &\left(\matrix{&&\cr &-E_8&\cr &&\cr}\right)&&&\cr
               &&\biggl(\skip U\skip\biggr)&&\cr
               &&&\biggl(\skip U\skip\biggr)&\cr
               &&&&\biggl(\skip U\skip\biggr)\cr}\right)\eqlabel{inters}$$
where $-E_8$ denotes the $8\times 8$ matrix given by minus the Cartan matrix of
the Lie algebra $E_8$ and $U$ is the "hyperbolic plane"
$$U\cong \left(\matrix{0&1\cr 1&0\cr}\right).$$
There is the natural embedding
$$\G_{3,19}\cong H_2(S,\IZ)\subset H_2(S,\IR)\cong \IR^{3,19}.$$
To "measure" the complex structure on our $K3$ surface we can use the periods
of the holomorphic 2-form $\O$
$$\varpi_i=\int_{e_i}\O.$$
Dividing $\O\in H^2(S,\IC)$ as $\O=x+iy$ where $x,y\in H^2(S,\IR)$ and noticing
that
$$\int_S\O\wedge\O=(x+iy).(x+iy)=(x.x-y.y)+2ix.y=0$$
and
$$\int_S\O\wedge\overline \O=(x+iy).(x-iy)=(x.x+y.y)=\int_S\Vert\O\Vert^2>O,$$
we conclude that the vectors $x$ and $y$ span a space-like 2-plane in
$H^2(S,\IR)$. If in addition we want to specify a Ricci-flat metric on $S$ we
need to choose a \K\ form, $J$, which represents another direction in
$H^2(S,\IR)$. This third direction is also space-like since
$$\int_SJ\wedge J={\rm Vol}(S)>0,$$
and it is perpendicular to $\O$ becuase the \K\ form is of type (1,1). Thus
$\O$ and $J$ together span a space-like 3-plane $\S$. As we rotate $\S$ in
$H^2(S,\IR)$ with respect to the fixed lattice $\G_{3,19}$ we obtain
inequivalent Ricci-flat metrics on the $K3$ surface. Thus the moduli space
of Ricci-flat metrics has the form of a Grassmanian of oriented space-like
3-planes in $\IR^{3,19}$ (modulo the effect of diffeomorphisms acting on the
lattice $H^2(S,\IZ)$) which locally isomorphic to
$${SO(3,19)\over SO(3)\times SO(19)}\times \IR_+,\eqlabel{modK}$$
where the $\IR_+$ factor describes the overall volume.
We still need to incorporate the $B$-field. Since $b_2(S)=22$ it will add 22
moduli to our 58 for a total of 80. Taking into account the holonomy of the
moduli space and the facts about conformal field theory we obtain a moduli
space which is locally of the form
$${SO(4,20)\over SO(4)\times SO(20)}\eqlabel{modKS}$$
(see~\cite{\Arev} for more detail).

Now to determine the form of the moduli space of sigma models on a $K3$
orbifold with two $E_8$ singularities we need to notice that in this case
the 3-plane $\S$ has to be perpendicular to all 16 roots of the two $E_8$'s
in~\eqref{inters}. Thus the moduli of the Ricci-flat metrics on the orbifold
will be given by a Grassmanian of a space-like 3-plane within the space
$\IR^{3,3}$, \ie locally isomorphic to
$${SO(3,3)\over SO(3)\times SO(3)}\times \IR_+.\eqlabel{modKo}$$
As to the $B$-field moduli, since we have shrunk 16 2-cycles to zero volume
we are left with only 6 of them for a total of $9+1+6=16$ moduli. Thus instead
of~\eqref{modKS} we get
$${SO(4,4)\over SO(4)\times SO(4)}.\eqlabel{modKSo}$$
\subsection{The heterotic "mirror" model}
To apply the {\bf c}-map we need to find the heterotic compactification
"mirror" to the one we have just described in the sense explained in \SS{1}.
The results of~\cite{\PR} tell us that the model we are looking for can be
constructed by taking a vector bundle with the structure group $E_8\times E_8$
(corresponding to the singularities of the orbifold) with 14 instantons in
one $E_8$ factor and 10 in the other. The resulting spectrum in four
dimensions consists of 3 vector
and 244 hypermultiplets.\Footnote{In six dimensions, the corresponding model
has 1 tensor, 244 hypermultiplets and no vectors so that the anomaly
cancellation condition holds as $H-V=244-0=244=273-29=273-29T$.}
Now it's the vector multiplet moduli space which is of interest to us.
Using~\eqref{ftree} and~\eqref{treemetric}we find that the tree level
prepotential reads
$$\ca F^{(0)}=-STU,\eqlabel{ftmir}$$
and the corresponding moduli space
$$\ca M_V^{(0)}={SU(1,1)\over U(1)}\times {SO(2,2)\over SO(2)\times SO(2)}.
\eqlabel{Mvtmir}$$
Note that~\eqref{Mvtmir} and~\eqref{modKSo} appear in the same row (second, for
$n=2$) of Table~\tabref{cmapsym} which means that the {\bf c}-map acting on
the tree level vector multiplet moduli space of our "mirror" model gives
precisely the classical moduli space for the hypermultiplets for the original
one thus providing some additional evidence for the conjecture of~\cite{\PR}.
To go beyond the classical moduli space for the hypermultiplets we have to use
the quantum corrected prepotential for vectors on the "mirror" side. To do that
we take the type IIA dual of the heterotic "mirror" model which in this case is
furnished by the \cyt\ described at the end of \SS{2}, namely the elliptic
fibration over $\IF_2$ with degenerate fibers of type I$_1$ and II only
(see~\cite{\Arev}). The exact prepotential for this \cym\ reads
$$\eqalign{\ca F=&-\bigl({4\over 3}t_1^3+t_1^2t_2+
2t_1^2t_3+t_1t_2t_3+t_1t_3^2\bigr)\cr
&+{1\over (2\p)^3}\sum_{d_1,d_2,d_3}n_{d_1,d_2,d_3}
{\rm Li}_3(e^{-2\p\sum_{A=1}^3 d_At_A})\cr}\eqlabel{exprepiia},$$
where some of the instanton numbers are given in Table~\tabref{instantons}
~\cite{\BKKM}. Using the map between type II and heterotic moduli~\cite{\L}:
$$\def\skip{\hskip15pt}
t_1=U,\skip t_2=S-T,\skip t_3=T-U,\eqlabel{map}$$
we can write the exact prepotential for the heterotic "mirror" model:
$$\ca F=-STU-{1\over 3}U^3+{1\over (2\p)^3}\sum_{l,m,k}n_{l+m+k,l,m+l}
{\rm Li}_3(e^{-2\p(lS+mT+kU)}).\eqlabel{exprephet}$$
\midinsert
$$\def\skip{\hskip10pt}
\def\extraspace{\omit{\vrule height3pt}&&&&&&&&\cr}
\vbox{\offinterlineskip\halign{\strut # height 12pt depth 6pt
&\hfil\skip \eightrm $#$\skip\hfil
&\hfil\skip \eightrm $#$\skip\hfil\vrule
&\hfil\skip \eightrm $#$\skip\hfil 
&\hfil\skip \eightrm $#$\skip\hfil\vrule
&\hfil\skip \eightrm $#$\skip\hfil
&\hfil\skip \eightrm $#$\skip\hfil\vrule
&\hfil\skip \eightrm $#$\skip\hfil
&\hfil\skip \eightrm $#$\skip\hfil\vrule\cr
\noalign{\hrule}
\omit{\vrule height5pt}&&&&&&&&\cr
\vrule&[0,0,1]&-2&[0,1,1]&-2&[0,1,2]&-4&[0,1,3]&-6\cr
\extraspace
\vrule&[0,1,4]&-8&[0,1,5]&-10&[0,2,3]&-6&[0,2,4]&-32\cr
\extraspace
\vrule&[1,0,0]&480&[1,0,1]&480&[1,1,1]&480&[1,1,2]&1440\cr
\extraspace
\vrule&[1,1,3]&2400&[1,1,4]&3360&[1,2,3]&2400&[2,0,0]&480\cr
\extraspace
\vrule&[2,0,2]&480&[2,2,2]&480&[3,0,0]&480&[3,0,3]&480\cr
\extraspace
\vrule&[4,0,4]&480&[5,0,0]&480&[6,0,0]&480&[0,1,0]&0\cr
\omit{\vrule height5pt}&&&&&&&&\cr
\noalign{\hrule}
}}
$$
\nobreak\tablecaption{instantons}{Some of the instanton numbers for rational
curves of multidegree $[d_1,d_2,d_3]$ for the \cym\ elliptically fibered over
$\IF_2$ with $(h_{11}, h_{21})=(3,243)$.}
\bigskip
\endinsert

\subsection{Corrections to the metric on the hypermultiplet moduli space}
Now that we have the expression~\eqref{exprephet} we can in principle write down
the $\a'$ corrections to the metric on the moduli space of hypermultiplets
of our original heterotic compactification in the regime of large $B_H$.
Let us rewrite~\eqref{exprephet} in a form that can
be directly substituted in~\eqref{dredLm}. We use the special coordinates
associated to the heterotic moduli.
$$\eqalign{\ca F=&-iz^1z^2z^3\cr
  &-{1\over 3}i(z^3)^3+{1\over (2\p)^3}\sum_{m,k}n_{m+k,0,m}{\rm Li}_3
  (e^{-2i\p(mz^2+kz^3)})\cr
  &+{1\over (2\p)^3}\sum_{l>0,m,k}n_{l+m+k,l,m+l}{\rm Li}_3
  (e^{-2i\p(lz^1+mz^2+kz^3)})\cr}\eqlabel{exprep}$$
The expression on the first line of~\eqref{exprep} is the tree level heterotic
prepotential for the "mirror" model and as we know gives rise,
via~\eqref{dredLm}, to the classical moduli space of sigma models on the
$K3$ orbifold with two $E_8$ singularities. Note that this is an indication
of the fact that under the {\bf c}-map the heterotic dilaton of the "mirror"
model maps to the hypermultiplet governing the overall area $A$ of the $K3$.
The piece of {\ca F} on the next line of~\eqref{exprep}
corresponds to one-loop correction to the "mirror" prepotential and, after the
{\bf c}-map (or, more precisely, the {\bf s}$_3$ transformation),
will give us the
$\a'$ corrections to the classical metric of the order of ${1\over A^p}$. And the
third line of that equation gives us the contribution to the "mirror" heterotic
prepotential arising non-perturbatively in $g_s$. On the hypermultiplet
side, it presumably will give us the $\a'$ corrections to the metric of the order
$e^{-A}$.

\section{disc}{Discussion}
If we represent an $N=2$ string vacuum in four dimensions
as a compactification of heterotic strings on a $K3\times T^2$ space
the hypermultiplet moduli space is free of corrections due to string
coupling. What remains to compute is the corrections due to finite size of
the compactification manifold. For that purpose one can use the fact that,
combining Heterotic/Type II
duality with Mirror Symmetry it is possible to find another heterotic
compactification which is in a sense "mirror" to the original one. Namely,
it produces a vacuum which is not equivalent to the one we have started with
but related to it in the following way. Suppose the type IIA dual to the
original heterotic model is determined by a \cym\ $M$.
Then the type IIA dual to the
"mirror" heterotic model is given by the mirror \cym\ $W$. This operation
produces different but not unrelated physics. The relation between the two
string vacua is that there is a well-defined operation, due to~\cite{\CFG},
transforming a special \K\ space into a quaternionic one that maps the moduli
space of vector multipltes of the "mirror" model to that of hypermultiplets
of the original one. This operation however is valid only at string tree level
in type II theories and hence cannot capture the quantum corrections which are
present in the geometry of hypermultiplets on type II side. As a result when
we apply it to heterotic models, although we start from the exact prepotential
for the "mirror" model the {\bf c}-map fails to produce the exact form of the
hypermultiplet moduli space metric, missing the part of the $\a'$ corrections
which would be mapped to the quantum corrections on the type II side under
the heterotic/type II duality. Thus all we can hope for is to obtain the $\a'$
corrections to the moduli space of our heterotic compactification in the
regime when the base of the elliptic fibration $S_H\rightarrow B_H$ is large
and we can ignore the quantum corrections on the type II side not captured
by the {\bf c}-map. Note that since all $\a'$ corrections disappear in the
limit of large overall volume of the $K3$, in order for
our results to make sense
we need to make the area of the ellptic fiber sufficiently small so that the
corrections resulting from the last two lines in \eqref{exprep} dominate the
corrections related to the type II dilaton and not captured by the
{\bf c}-map.  

Specifically, heterotic compactifications involving small instantons
provide us with models with a few hypermultiplet moduli whose "mirrors"
have correspondingly low-dimensional vector multiplet moduli spaces for which
exact results have in fact been obtained previously. We have found, in one
particular example, that the tree-level vector multiplet moduli space of the
"mirror" produces, under the {\bf c}-map, precisely the classical moduli space
of the hypermultiplets. Thus the corrections which on the "mirror" side are due
to finite string coupling are responsible to the corrections to the classical
hypermultiplet moduli space which are attributed to finite size of the $K3$.
We find that in a sense the $g_s$ expansion on the "mirror" maps into the $\a'$
expansion on the original heterotic model.\Footnote{In the limit of large $B_H$}
We have not
written down the resulting metric explicitly mainly because the expressions
seem to be quite complicated. Let us note however that as it is the case
with the vector multiplet moduli spaces the dual-quaternionic manifolds
which obtain are fully specified by a single holomorphic function
we have written down explicitly. As we have said already the biggest omission
is our failure to compute the $\a'$ corrections away from the large $B_H$ limit.
In order to do that one would have to find the "quantum" {\bf c}-map which in
effect amounts to computing the quantum corrections to the moduli space of
hypermultiplets on the type II side. Some interesting results in this direction
were obtained in
\Ref\BBS{K. Becker, M. Becker and A. Strominger, \npb{456} (1995) 130,
hep-th/9507158.}\ and subsequent progress made in
\Ref\OV{H. Ooguri, C. Vafa, Phys. Rev. Lett. {\bf 77} (1996) 3296,
hep-th/9608079.}\ and
\Ref\Sh{A. Strominger, \plb{421} (1998) 139, hep-th/9706195.}.
Another point worth mentioning is that in the
present paper we were dealing with the local form of the moduli space only
leaving completely out the questions of discrete identifications due to
dualities. We leave these issues for future work.

\vskip5pt
\noindent {\bf Acknowledgements}
\vskip5pt
\noindent It's a pleasure to thank P.~Candelas, X.~de~la~Ossa and S.~T.~Yau
for helpful discussions. I am grateful to E.~Witten and P.~Mayr for pointing
out an error in the original version of the paper.
This work was supported by the Department of Energy grant.

\newpage
\immediate\closeout\referencewrite\referenceopenfalse
\line{\bf\hfil References\hfil}\bigskip\parindent=0pt\input referenc.texauxil
\bye